\documentclass[]{aastex631}

\graphicspath{{./}{figures/}}

\begin{document}

\title{QLP Data Release Notes 002: Improved Detrending Algorithm}

\correspondingauthor{Michelle Kunimoto}
\email{mkuni@mit.edu}

\author[0000-0001-9269-8060]{Michelle Kunimoto}
\affiliation{Department of Physics and Kavli Institute for Astrophysics and Space Research, Massachusetts Institute of Technology, Cambridge, MA 02139}

\author[0000-0002-5308-8603]{Evan Tey}
\affiliation{Department of Physics and Kavli Institute for Astrophysics and Space Research, Massachusetts Institute of Technology, Cambridge, MA 02139}

\author[0000-0003-0241-2757]{Willie Fong}
\affiliation{Department of Physics and Kavli Institute for Astrophysics and Space Research, Massachusetts Institute of Technology, Cambridge, MA 02139}

\author[0000-0002-2135-9018]{Katharine Hesse}
\affiliation{Department of Physics and Kavli Institute for Astrophysics and Space Research, Massachusetts Institute of Technology, Cambridge, MA 02139}

\author[0000-0002-1836-3120]{Avi Shporer}
\affiliation{Department of Physics and Kavli Institute for Astrophysics and Space Research, Massachusetts Institute of Technology, Cambridge, MA 02139}

\begin{abstract}
Light curves feature many kinds of variability, including instrumental systematics, intrinsic stellar variability such as pulsations, and flux changes caused by transiting exoplanets or eclipsing binary stars. Detrending is a key pre-planet-search data processing step that aims to remove variability not due to transits. This data release note describes improvements to the Quick-Look Pipeline's detrending algorithm via the inclusion of quaternion data to remove short-timescale systematics. We describe updates to our procedure, intermediate data products outputted by the algorithm, and improvements to light curve precision.
\end{abstract}


\section{Updates}

\subsection{Detrending Algorithm}

Since the start of the Transiting Exoplanet Survey Satellite mission \cite[TESS;][]{Ricker2015}, the Quick-Look Pipeline (QLP) has detrended light curves from each TESS orbit following the basis spline method of \citet{VanderburgJohnson2014}. In short, each ``raw'' light curve is divided by a basis spline fit to produce a ``detrended" light curve. The spacing of the spline break points is chosen by minimizing the Bayesian Information Criterion (BIC), imposing a minimum allowed spacing of 0.3 days.

Our adoption of a new algorithm is primarily motivated by a desire to improve the removal of instrumental systematics. Towards this end, in the new detrending method we use camera quaternion time series in addition to the basis splines already in use \cite[similar to ][]{toi-396, toi-1130}.

The camera quaternion data are three-dimensional time series data that describe spacecraft pointing every two seconds.\footnote{Quaternion data can be found online at \url{https://archive.stsci.edu/missions/tess/engineering/}.} These data are then binned down to the observing cadence (200s for the second TESS Extended Mission) and we compute the mean, standard deviation, and skew within each bin to get nine binned quaternion time series (for a particular camera). With these, we compute the products between series of the same statistic type. For example, we multiply the binned means from the first and second dimensions of the quaternions together to produce a new series. We do the same for the first and third dimensions as well as the second and third dimensions before moving onto standard deviation products. We then compute the squares of all previously computed series. 

Combined, we have a total of 36 feature series with relevant systematics information, particularly on shorter timescales. We normalize all feature series to have zero mean and unit variance. We then jointly fit these feature series along with basis splines to the star's raw light curve with an iterative linear least-squares fit, removing 3-$\sigma$ outliers after each iteration. Like before, we choose a basis spline spacing that minimizes BIC with a minimum allowed spacing of 0.3 days. The final best fit is divided out of the raw light curve to produce the detrended orbit light curve. 

\begin{figure*}
    \centering
    \includegraphics[width=\textwidth]{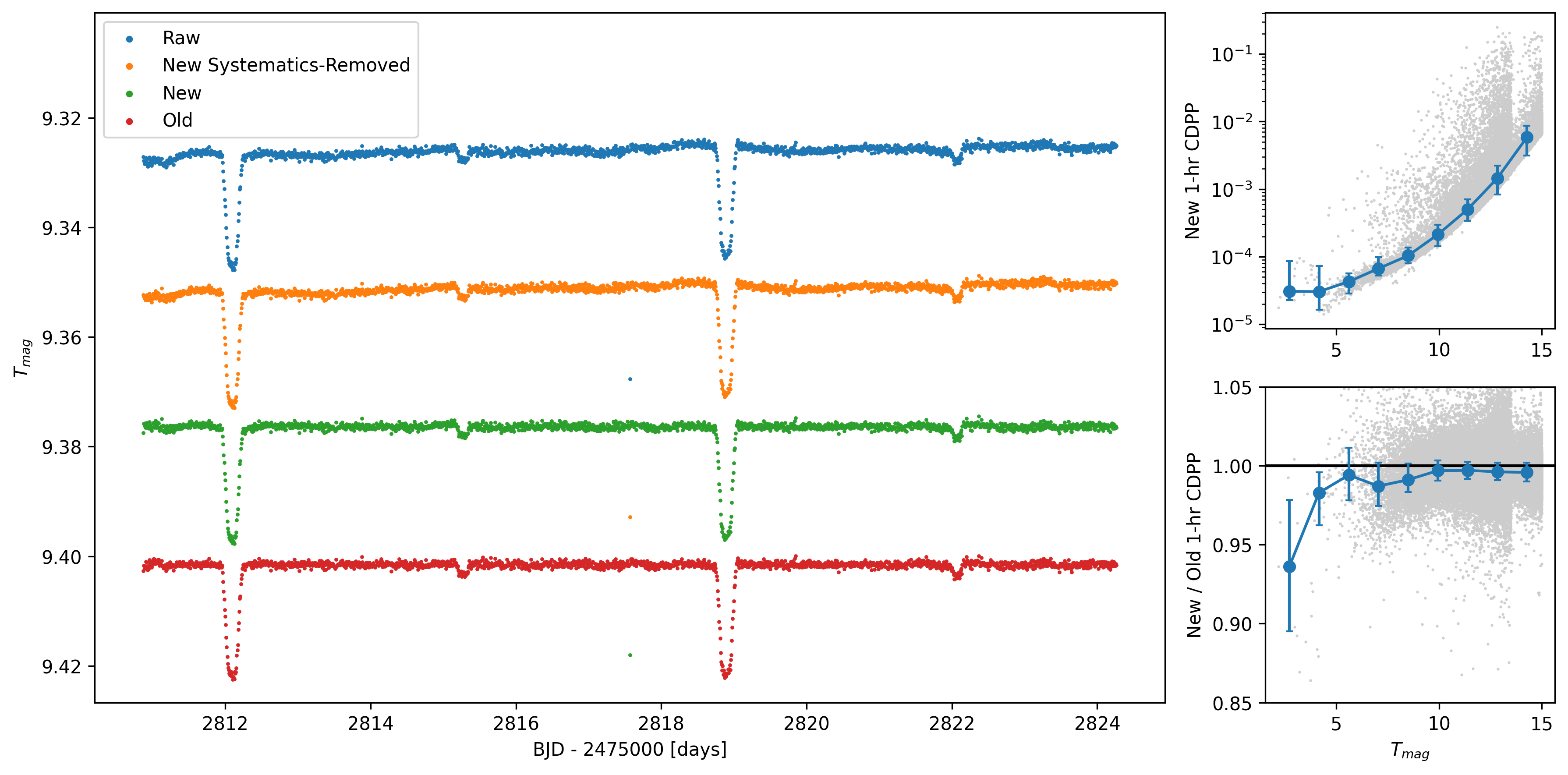}
    \caption{Left: Raw, new systematics-removed, new detrended, and old detrended light curves for TIC 237117591 in Orbit 118. Raw light curve extraction is described in \citet{Huang2020}. The new detrending method jointly fits basis splines and quaternion data, then removes the fit from the raw light curve. The quaternion data help decorrelate short-timescale systematics from the light curve where the old method only used basis splines to remove long-timescale trends. A byproduct of the new method is the systematics-removed light curve which removes the short-timescale systematics while leaving astrophysical phenomena like transits and intrinsic variability in the light curve. Top right: 1-hr CDPP vs $T$ mag for the new detrending method with 10th and 90th percentiles marked with error bars. Bottom right: Ratio of new to old 1-hr CDPP vs $T$ mag with 25th and 75th percentiles marked with error bars. Overall, the new detrending method reduces noise across all $T$ mag levels -- especially reducing noise for brighter stars.}
    \label{fig}
\end{figure*}

Figure \ref{fig} shows the raw light curve with old and new detrended light curves from Orbit 118 (2022 August 19 -- 31; First Extended Mission) for TIC 237117591 ($T = 9.33$ mag), as well as an intermediate version of the light curve with the systematics removed but astrophysical variability left intact. We also show the 1-hour Combined Differential Photometric Precision (CDPP) of the new method relative to the old method on Camera 4 targets in Orbit 118. In particular, we see improvements across all TESS magnitudes, especially for bright stars ($T \lesssim 7$ mag) for which CDPP improved on the order of several percent.

Starting in Sector 56, the new detrending method has become a regular part of processing and will produce the detrended light curves provided in our High-Level Science Products (HLSPs).\footnote{\url{https://archive.stsci.edu/hlsp/qlp}} Our HLSPs will also include the light curves with only systematics removed. These systematics-corrected light curves will improve the suitability of QLP to studies of astrophysical variability, such as pulsations, oscillations, and contact binaries. We have future plans to reprocess all older orbits (118 and earlier) with the new detrending method.

\section{Summary}

Effective Sector 56 and onward, QLP underwent the following changes to improve light curve detrending for astrophysical variability studies and planet searches:

\begin{itemize}
    \item The adoption of a new detrending algorithm that decorrelates short-timescale systematics via quaternion data from each star's raw light curve.
    \item The production of intermediate light curves that preserve astrophysical variability, but remove instrumental systematics.
\end{itemize}

\section{Acknowledgements}
These data release notes provide processing updates from the Quick Look Pipeline (QLP) at the TESS Science Office (TSO) at MIT. QLP extracts and detrends light curves from TESS Full-Frame Images (FFIs), searches light curves for transits, and produces vetting reports for promising planet candidates. The full QLP procedure is described in \citet{Huang2020, Kunimoto2021}. This work makes use of FFIs calibrated by TESS Image CAlibrator \citep[TICA;][]{TICA}, which are also available as High-Level Science Products (HLSPs) stored on the Mikulski Archive for Space Telescopes (MAST). Funding for the TESS mission is provided by NASA's Science Mission Directorate.

\bibliography{refs}

\end{document}